\documentclass[usenatbib]{mn2e}
\usepackage{graphicx}
\usepackage{fixltx2e}
\usepackage{textcomp}

\newcommand{\Msun}{\rm{\,M_\odot}}
\newcommand{\Msunh}{\rm{\,M_\odot/h}}
\newcommand{\Mpch}{\rm{\,Mpc/h}}

\newcommand{\Rv}{R_{\rm{178}}}
\newcommand{\Cs}{C_{\rm{178}}}
\newcommand{\Vc}{v_{\rm{circ}}}

\newcommand{\dd}{\mathrm{d}}
\newcommand{\kpc}{\mathrm{kpc}}

\begin{document}
\title{Virialization of high redshift dark matter haloes}

\author[Davis, D'Aloisio \& Natarajan]{Andrew J. Davis,$^1$ Anson D'Aloisio,$^2$ and Priyamvada Natarajan$^{1,2}$ \\
$^1$Department of Astronomy, Yale University, P.O. Box 208101, New Haven, CT 06520-8101, USA\\
$^2$Department of Physics, Yale University, P.O. Box 208120, New Haven, CT 06520-8120, USA}

\maketitle
\begin{abstract}
We present results of a study of the virial state of high redshift
dark matter haloes in an N-body simulation. We find that the majority
of collapsed, bound haloes are not virialized at any redshift slice in
our study ($z=15-6$) and have excess kinetic energy. At these
redshifts, merging is still rampant and the haloes cannot strictly be
treated as isolated systems. To assess if this excess kinetic energy
arises from the environment, we include the surface pressure term in
the virial equation explicitly and relax the assumption that the
density at the halo boundary is zero. Upon inclusion of the surface
term, we find that the haloes are much closer to virialization,
however, they still have some excess kinetic energy.  We report trends
of the virial ratio including the extra surface term with three key
halo properties: spin, environment, and concentration.  We find that
haloes with closer neighbors depart more from virialization, and that
haloes with larger spin parameters do as well.  We conclude that
except at the lowest masses ($M < 10^6 \Msun$), dark matter haloes at
high redshift are not fully virialized. This finding has interesting
implications for galaxy formation at these high redshifts, as the
excess kinetic energy will impact the subsequent collapse of baryons
and the formation of the first disks and/or baryonic structures.
\end{abstract}


\section{Introduction}
\label{sec:intro}

In the standard $\Lambda$ cold dark matter (LCDM) model, small
Gaussian perturbations in the dark matter density field at early times
provide the seeds for the formation of structure in the Universe. The
dark matter gravitationally collapses, and forms bound structures that
eventually relax into a state of virial equilibrium.  The number
density of collapsed dark matter haloes at a given mass and epoch, the
mass function, of these relaxed haloes provides powerful constraints
on the parameters of the LCDM model \citep[see
e.g. ][]{Haiman01,Cunha10}. Various analytical methods
\citep[e.g. ][]{PS, Bond91, ST} also predict the halo mass function
for collapsed, bound, and \emph{virialized} haloes in LCDM. However,
simulations at high redshift ($z > 1$) have found that the majority of
collapsed, bound haloes are not in virial equilibrium \citep{JCH01,
Hetz06, Davis10}.  Thus, there seems to be a mis-match with
simulations; they find mass functions (which assume the haloes are
virialized) that match the analytic predictions, and yet the detailed
structure of the haloes shows that they are not virialized.  In this
paper, we explore in detail the virialization state of dark matter
haloes at high redshift in order to understand this discrepancy.

For an isolated collapsed, bound dark matter halo in equilibrium, the scalar 
virial theorem,
\begin{equation}
2K + U = 0,
\label{EQ:scalarVE}
\end{equation}
provides a simple relationship between the halo's total kinetic ($K$)
and potential ($U$) energies. In LCDM, dark matter haloes are expected
to reach virial equilibrium rapidly upon collapse when they detach
from the Hubble flow. The timescale for virialization is of the order
of the dynamical time, which for a dark matter halo may be estimated
as $t_r \approx \Rv/\Vc$, where $\Rv$ is the virial radius and $\Vc$ the
circular velocity, $\Vc = \sqrt{GM/\Rv}$.   For a $10^7 \Msun$
halo at $z=6$ this is roughly $1 \times 10^8~ \mathrm{yrs}$, or one percent
of the Hubble time at that redshift. Therefore, despite rapid merging
activity these haloes have had sufficient time to reach virial
equilibrium, but do not appear to do so in the simulations.

Here we explore the energy budget of these haloes, to determine why
simulated haloes at high redshift are apparently out of virial
equilibrium.  There are several possibilities which may explain this
finding.  In this paper, we probe this issue by relaxing two
assumptions typically made when applying the virial theorem. First, we
include the non-negligible contributions of the environment to the
halo's gravitational potential and secondly, we do not truncate the
density profile of the halo at the virial radius.  While these two
assumptions are valid for isolated haloes typical of the local
Universe, at high redshift when the Universe was denser, these
assumptions are incorrect. In addition to these two assumptions, it is
possible that systematic errors from the halo finding algorithm may
bias measurements of halo virialization.  Finally, we explore the possibility
that the departure from virialization is correlated with key halo
properties. The organization of this paper is as follows: we derive
the virial equation from the momentum equation in section
\ref{sec:ve}, explicitly retaining the surface terms that are usually
neglected. In section \ref{sec:sim} present simulation results and the
correlations between halo energetics and key halo structural
properties, namely, halo spin, concentration, and local environment.
We conclude in section \ref{sec:discussion} with a discussion of our
findings and their implications.


\section{The Virial Equation}
\label{sec:ve}

Here we derive the virial theorem for dark matter haloes
including boundary effects and an external gravitational potential.  
An isolated system of collisionless particles in
equilibrium satisfies the scalar virial theorem (see Equation
\ref{EQ:scalarVE}). This result is derived under the assumption that
the mass density approaches zero at large radii for a finite mass
distribution.  In this work, our systems of interest are high
redshift dark matter haloes, which form within a cosmic web of sheets,
voids, and filaments. We therefore cannot assume that the mass
density approaches zero at the boundary of these systems. We also consider 
the potential due to nearby particles exterior to the haloes.\footnote{In
what follows, we use the Einstein summation convention where repeated
indices correspond to summations.}
 
The virial theorem is obtained from the conservation of momentum
for a dark matter halo modeled as a collisionless fluid.  Following
\citet{BTbook}, the corresponding fluid equation is multiplied by
$x_k$ and integrated over the volume $V$, with the caveat that the
surface terms resulting from integration by parts do not vanish.  In
this case, the tensor virial theorem corresponding to a collisionless
fluid is given by,

 \begin{equation}
 \frac{\dd^2 I_{ij}}{\dd t^2} = 2 K_{ij} + W_{ij}- \oint{ \rho x_i v_j v_k n_k \dd S}  - \frac{1}{2} \frac{\dd}{\dd t}\oint{ \rho x_i x_j v_k n_k \dd S},
 \label{EQ:tensorvir}
 \end{equation}
 where $I_{ij} \equiv \int_V{\rho x_i x_j \dd^3 \vec{x}}$ is the
 moment of inertia tensor, $K_{ij} \equiv \frac{1}{2}\int_V{\rho
 v_iv_j \dd^3\vec{x}}$ is the kinetic energy tensor, $W_{ij} \equiv -
 \int_V{\rho x_i\partial \Phi/\partial x_j \dd^3 \vec{x}}$ is the
 potential energy tensor.  We denote the mass density, position,
 average velocity, and gravitational potential by $\rho$, $\vec{x}$,
 $\vec{v}$, and $\Phi$ respectively.  The vector $\vec{n}$ is the
 outward-pointing unit normal to the surface $S$ that encloses the
 volume $V$.  The third term on the right of equation
 (\ref{EQ:tensorvir}) corresponds to the kinetic stresses on the halo
 boundary.  The last term is the time derivative of the moment of
 inertia flux through the surface \citep{BP06}. The scalar virial theorem 
is obtained by taking the trace of equation (\ref{EQ:tensorvir}):
 
\begin{equation}
 \frac{\dd^2 I}{\dd t^2} = 2 K + W- \oint{ \rho \vec{x}\cdot \vec{v}~\vec{v}\cdot \dd \vec{S}}  - \frac{1}{2} \frac{\dd}{\dd t}\oint{ \rho x^2~\vec{v}\cdot \dd \vec{S}}. 
 \label{EQ:scalarvir}
 \end{equation} 
 Following \citet{BTbook} and \citet{BP06}, the potential energy term
 may be broken up into contributions from particles inside and outside
 of the halo, $\Phi = \Phi_{\mathrm{int}} + \Phi_{\mathrm{ext}}$, so
 that
 
 \begin{equation}
 W = -\frac{1}{2} \int_V{\rho~\Phi_{\mathrm{int}} \dd V} -\int_V{\rho~\vec{x}\cdot \frac{\partial{\Phi_{\mathrm{ext}}}}{\partial{\vec{x}}} \dd V}.
 \end{equation}
 Note that the first term is the gravitational potential energy of the
 halo, which we denote as $U$ from here on.  In the case of a
 spherical volume, the last term in equation (\ref{EQ:scalarvir}) is
 $R^2 \ddot{M}/2$, where $R$ is the radius and $M$ is the mass enclosed 
within the volume.  Assuming that this term is negligible for a steady-state
 system, we obtain
\begin{eqnarray}
0 &= &2 K + U + U_{\mathrm{ext}}- E_s,  \label{EQ:vir}\\
 U_{\mathrm{ext}} & \equiv & -  \int_V{\rho~\vec{x}\cdot \frac{\partial{\Phi_{\mathrm{ext}}}}{\partial{\vec{x}}} \dd V}, 
 \label{EQ:Uext} \\
 E_s & \equiv & \oint{ \rho \vec{x}\cdot \vec{v}~\vec{v}\cdot \dd \vec{S}}, \label{EQ:Es}
 \end{eqnarray}  
 for a halo in virial equilibrium.  $U_{\mathrm{ext}}$ and $E_s$
 represent corrections to the standard virial relation
 and can be explicitly calculated for simulated dark matter haloes.


\section{Simulation Results}
\label{sec:sim}
Using GADGET-2 \citep{GADGET}, we ran a dark matter only simulation
with the WMAP5 cosmological parameters \citep[\{$\Omega_M,
\Omega_\Lambda, \Omega_b, h, n, \sigma_8$\} = \{0.258, 0.742, 0.044,
0.719, 0.963, 0.796\},][]{wmap5} from $z \approx 100$ to $z = 6$ (see
\citet{Davis10} for full details).  Our dark matter particle mass is
$m = 1.0 \times 10^4 \Msunh$, which sets a comoving box size at
$2.46 \Mpch$, with $512^3$ dark matter particles.  We use the HOP
algorithm to identify collapsed dark matter haloes \citep{HOP}.  HOP 
first calculates a density for each particle by smoothing over its nearest 64 neighbors using a cubic spline kernel.
 It then groups particles with their densest neighbor, and density
thresholds are used to ensure that haloes are not being over-counted
as sub-haloes within a larger halo.  We choose density thresholds
 to match the high redshift mass function described in
\citet{reed07}.  Lastly, we remove unbound particles from each halo.  

For each halo, we calculate the total potential energy, $U$, using a
direct summation over the particles assigned to the halo:
\begin{equation}
U = - \sum_{i=1}^{N-1} \sum_{j=i+1}^{N} \frac{Gm_i m_j}{r_{ij}}, \label{eqn:pot}
\end{equation}
 where $r_{ij}$ is the separation between particles $i$ and $j$, $G$
 is Newton's gravitational constant, and $m$ the particle mass.  The
 total kinetic energy, $T$ is found by summing over the particle's
 individual kinetic energies:
 \begin{equation}
 T = \frac{1}{2} \sum_{i=1}^N m_i \vec{v}_i^2.
 \end{equation}
 
\begin{figure}
\begin{center}
\includegraphics[scale=0.3,angle=90]{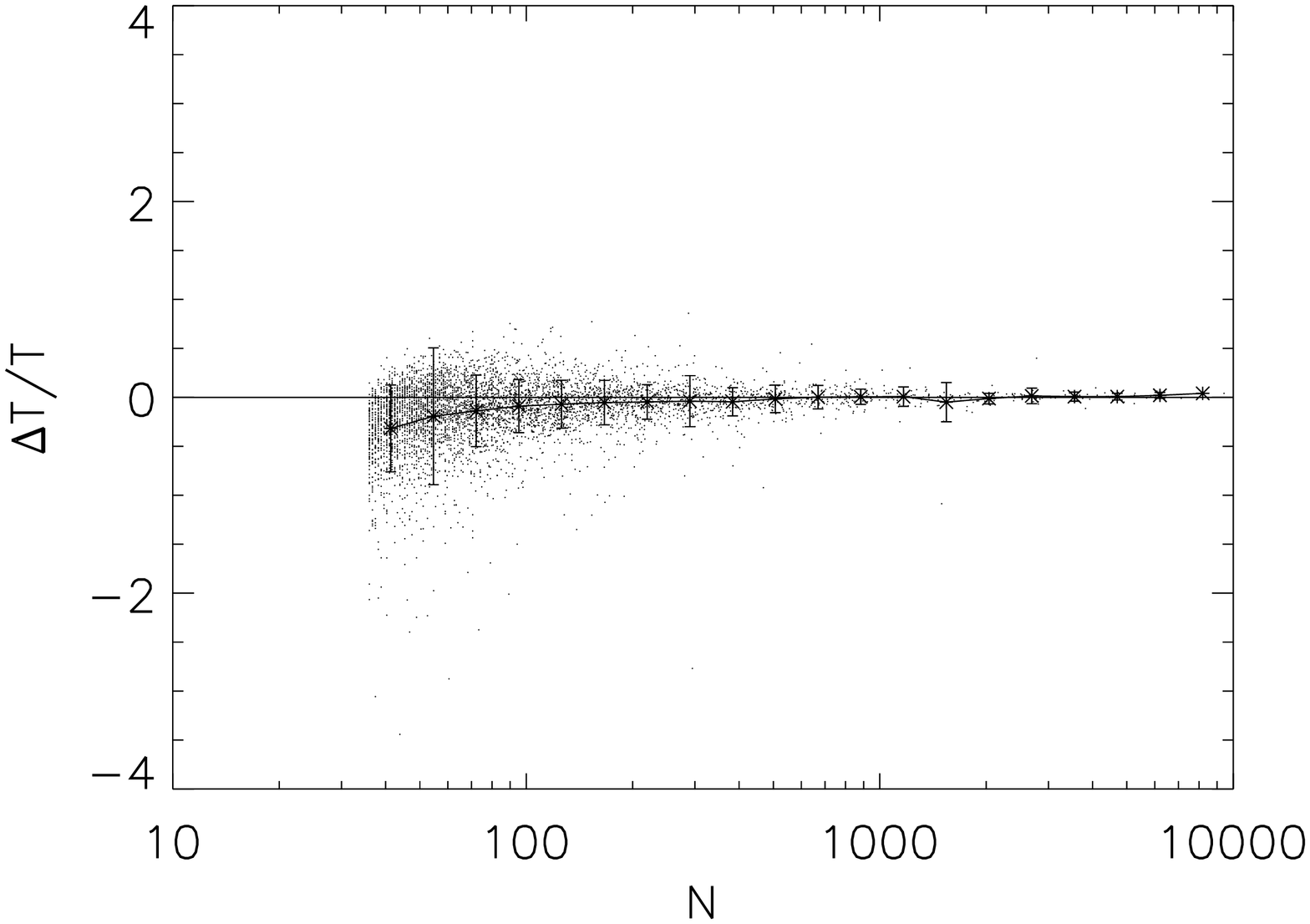}
\includegraphics[scale=0.3, angle=90]{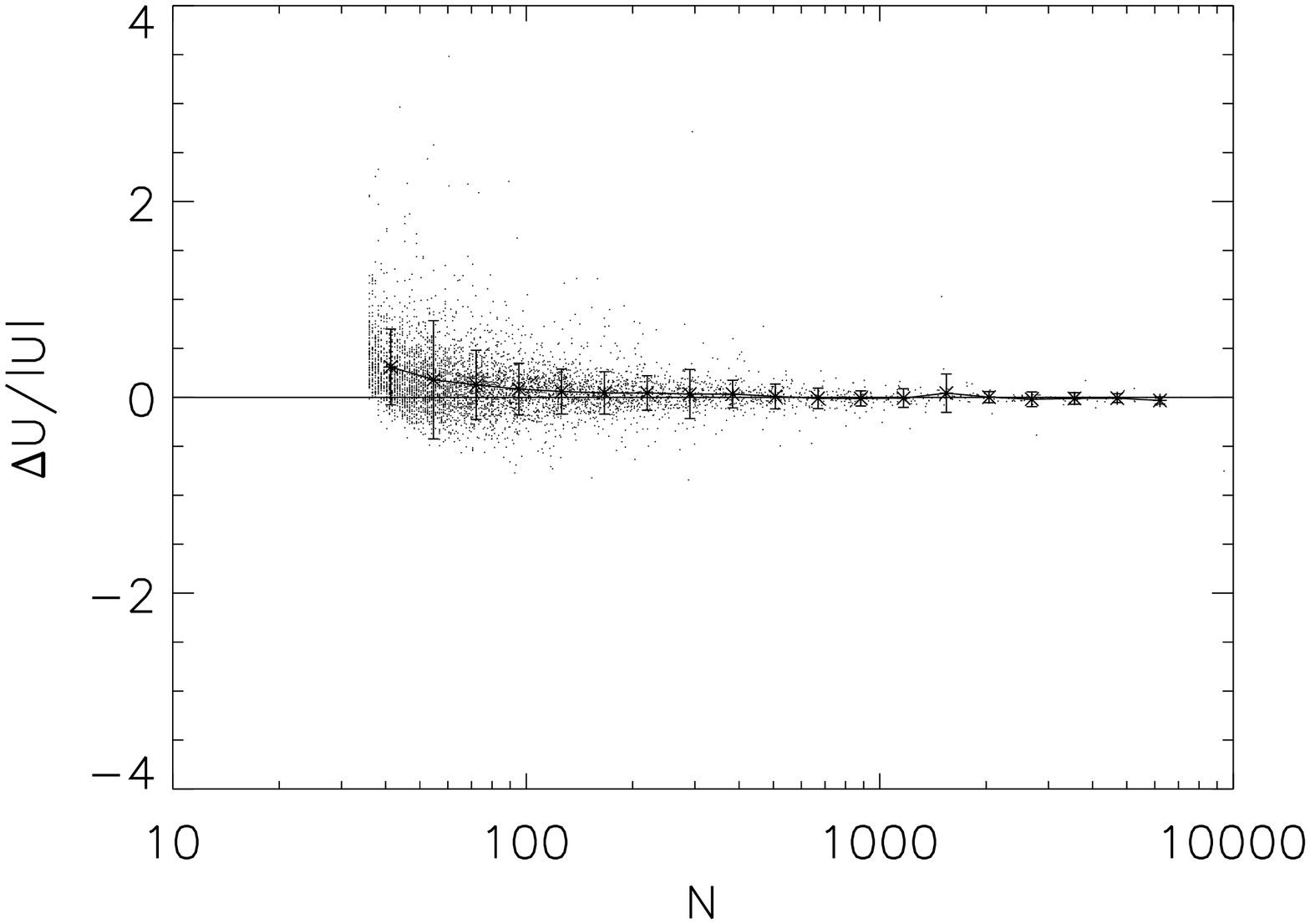}
\caption{Fractional difference of $T$ (top) and $U$ (bottom) as a function of the number of particles in the low resolution halo.  The error bars represent the $1 \sigma$ deviation about the mean.  We find that at least $300$ particles are required to accurately measure the kinetic and potential energies in our sample.}
\label{Fig:convergence}
\end{center}
\end{figure}

In measuring $T$ and $U$, we must ensure that we have enough 
particles in our halo sample to accurately measure the energies.  We addressed this question
in \citet{Davis10}, and summarize the relevant points here.  We re-ran our simulation with one eighth 
as many particles ($256^3$ total) and with eight times as many particles ($1024^3$ total).
This allows us to compare directly individual halos with different resolutions.
We used the halo centre of mass to cross-matched the halo catalogues.   We show in Figure
 \ref{Fig:convergence} the change in $T$ and $U$ for the $512^3$ and $256^3$ runs as a 
 function of the number of particles in the low resolution halo.  We find that at least $300$ 
 particles are required to accurately measure the kinetic and potential energies, and restrict our halo 
 sample to halos with at least this many particles, corresponding to a halo mass of $M = 3 \times 10^6 \Msunh$.
 
Having measured the kinetic and potential energies of our haloes, we can define the virial ratio, 
\begin{equation}
\beta = 2T/U + 1,
\end{equation}
which for fully virialized haloes should be zero.  However, as our
measurements of $T$ and $U$ are instantaneous quantities, and not time
averaged, we expect a range of values for $\beta$, with a mean value of zero.  There are various
cuts on $\beta$ used in the literature to select strictly virialized
haloes: \citet{Shaw06} use $\beta > -0.2$, \citet{Bett07} use $|\beta|
< 0.5$, and \citet{Neto07} use $\beta < 0.35$.  As noted in
\citet{Davis10} we find that our haloes do not have
$\left<\beta\right> \approx 0$, but are offset to high values of
kinetic energy such that $\left<\beta\right> < 0$ for all redshifts,
and haloes are further from virialization at higher redshifts.  One
possibility is that the extra terms in the virial equation
($U_{\mathrm{ext}}$ and $E_s$) are not negligible at higher redshifts, where large
amounts of infalling material contribute to the terms.  As the
Universe expands, we expect a smaller contribution from these extra terms.  Our
findings follow the general trend reported in \citet{Hetz06} at lower
redshift, though their fitting function cannot be extended to our
redshift range.

\begin{figure}
\begin{center}
\includegraphics[scale=0.3,angle=90]{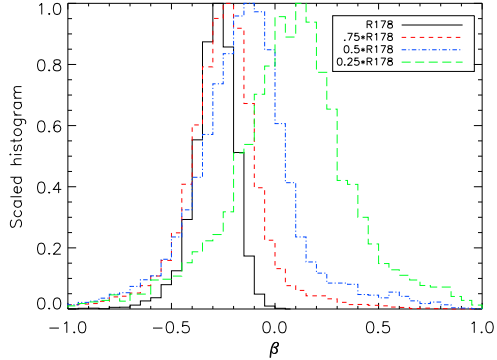}
\caption{Scaled histogram of $\beta$ evaluated at four different radii.  We find that the cores of our haloes are
in virial equilibrium, while the overall halo is not.}
\label{Fig:core}
\end{center}
\end{figure}

The virialization process is expected to happen from the inside out.  Thus,
it should be possible to identify a virialized core.  To do this, we calculated $T$ and $U$ only 
for particles inside three smaller 
radii: $0.75 \Rv, 0.5 \Rv,$ and $0.25 \Rv.$   For the choice of boundary at $0.25\Rv$, we do find 
virialized cores, as expected.  We
show in Figure \ref{Fig:core} histograms of $\beta$ for these three inner radii as well as the histogram of
$\beta$ for the entire halo.  We find that for smaller radii, the mean value of $\beta$
shifts towards $0$, as expected for a virialized object.  Thus we conclude that the halo cores of our
sample are virialized, while the entire halo is not. 

The departure from virial equilibrium motivated our interest in the extra terms in the virial
equation.  We calculate the contribution to the potential energy from the
external mass distribution, $U_{\mathrm{ext}}$ (equation \ref{EQ:Uext}). 
Converting the volume integral into a sum over particles, we find
\begin{equation}
U_{\mathrm{ext}}  = - \sum_i^{N} m_i \; \vec{x}_i \cdot \vec{a}_i 
\label{eqn:sumW},
\end{equation}
where $N$ is the total number of particles in the halo (see equation
\ref{eqn:sumW}).  To calculate the gravitational acceleration
$\mathbf{a}$ acting on particle $i$ we sum over all particles, $j$,
within $10 \Rv$ which are not part of the halo itself:
\begin{equation}
\mathbf{a}_i = - \sum_j^{N_{\rm{ext}}} \frac{G m_j \mathbf{r}_{ij}}{r_{ij}^3}. 
\end{equation}
However, after calculating this term, we find that it is negligible
compared to $U$ (of order $1\%$).  Thus, even at higher
redshifts where it may be expected that external matter will affect
the total gravitational term in the virial equation, we find that it
provides only a small contribution to the total energy.

We next examine the term in equation \ref{EQ:Es}, $E_s$.  Following \citet{Shaw06} ,
we select halo particles between $0.8 \Rv$ and $\Rv$, where $\Rv$ is the
radius which encloses a mean overdensity of $178 \rho_{\rm{crit}}$.  
This defines the surface over which we perform the summation.  Then, approximating
the surface integral as a summation over particles in this shell, we
calculate $E_s$ as:
\begin{equation}
E_s = \frac{R_m ~ m }{\Delta r} \sum_i{v_{i,r}^2}
\label{eqn:betaprime}
\end{equation}
where $R_m$ is the mean radius of the $N_s$ particles in the shell of thickness $\Delta r = 0.2 \Rv$.
Each particle, $i$, has a radial velocity $v_{i,r} = \vec{r}_i \cdot
\vec{v}_i / |\vec{r}_i|$.  We note that
 our $E_s$ term is similar to that of \citet{Shaw06}, but not identical to their surface 
 pressure term.  Here we included only the contribution from the velocity 
 component normal to the surface.  This may be important if the velocities of the 
 particles falling into the halo are on predominantly radial orbits, such as falling 
 into the halo along a filament, and would then have a larger surface term than when simply 
 calculating the total pressure (which is proportional to $|\vec{v}|^2/3$).  We find that at least $30$ particles are
required in the shell to correctly resolve and calculate $E_s$; haloes
with fewer particles have strongly biased (low) values of $E_s$.  We
therefore only include haloes with more than 30 particles in this shell for the remainder of
this work. 

We point out that our halos are not spherical -- there is no guarantee that the density cuts 
used by the HOP algorithm yields spherical halos.  However, in our calculation of $E_s$, we
have chosen a spherical surface over which to measure this term.  Thus, we wish to test whether
there is any bias in the value of $E_s$ arising from the real shapes of haloes.  To measure halo shapes,
we use the same method as in \citet{Davis10}.  First, we calculate the normalized moment of inertia
tensor.  The ratios of the eigenvalues ($ a > b > c$) of this tensor can be used to define a halo's sphericity 
($s = c/a$) and triaxiality ($t = (a^2-c^2)/(a^2-b^2)$).  We found, however, no trend between shape and 
$E_s$ or $\beta$.  Thus, we conclude that our choice of using a spherical shell rather than following the
outer boundary of the HOP halo does not bias our results.

\begin{figure}
\begin{center}
\includegraphics[scale=0.6]{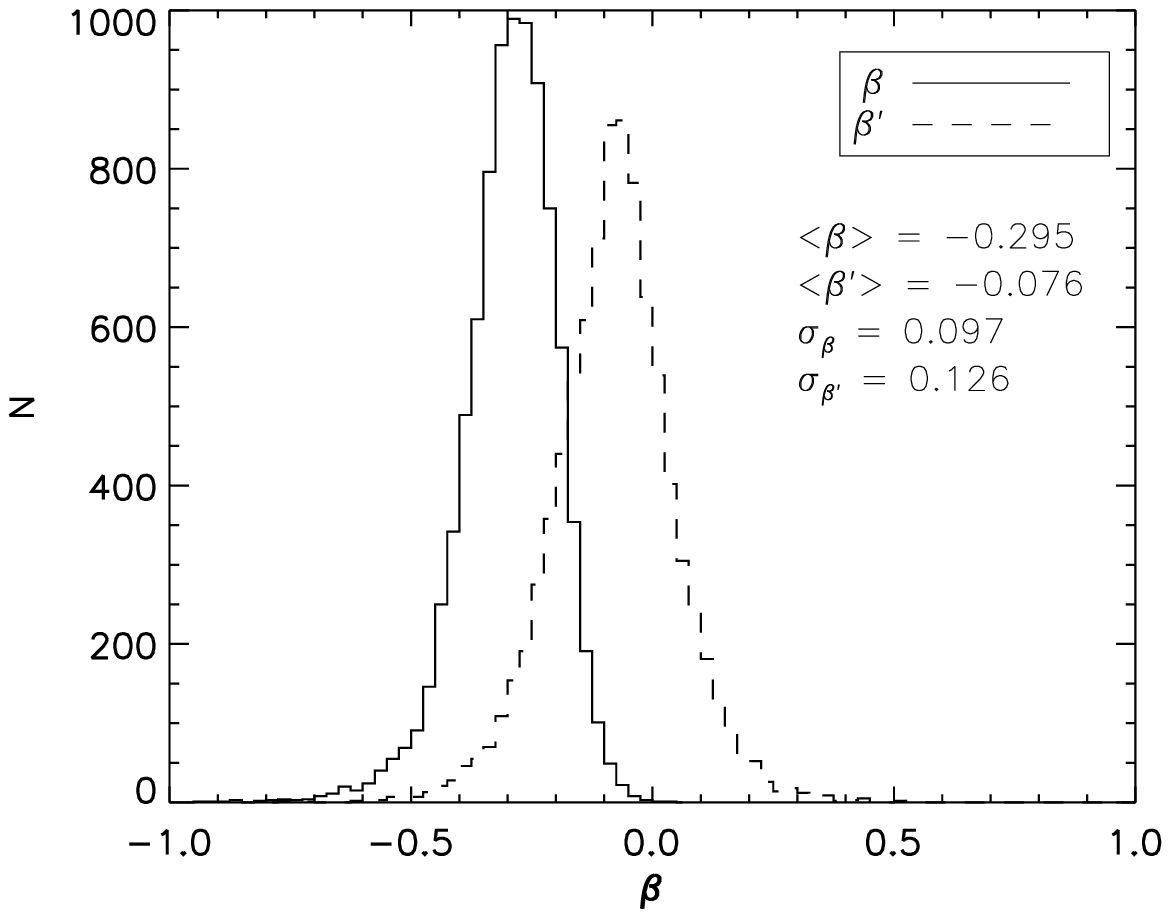}
\includegraphics[scale=0.6]{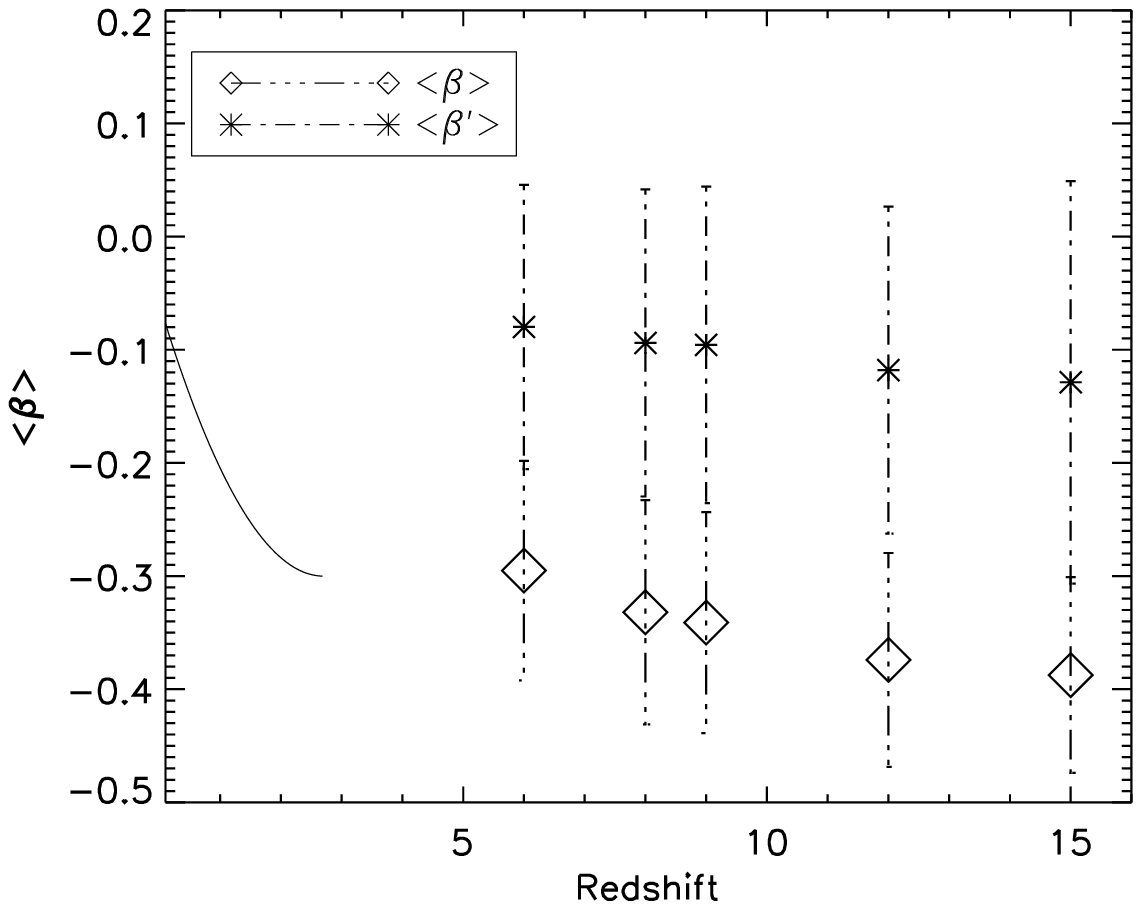}
\caption{Top panel: histograms of the virial ratio, $\beta$, and the
corrected virial ratio, $\beta'$, at $z=6$.  The mean and standard
deviation of each distribution is also shown.  Bottom panel: mean
virial ratio, $\left<\beta \right>$ (diamonds), and mean corrected
virial ratio, $\left< \beta' \right>$ (asterisks), as a function of
redshift.  The solid curve is the fitted polynomial for
$\left<\beta\right>$ provided by \citet{Hetz06} for redshifts $z < 3$.
The error bars represent the $1 \sigma$ deviation about the mean.}
\label{Fig1}
\end{center}
\end{figure}

Having found that $U_{\mathrm{ext}}$ is negligible, we now include only
 the surface pressure term, $E_s$, in our calculation of a corrected virial
 ratio, $\beta'$, where: 
\begin{equation}
\beta' = (2T - E_s)/U+1,
\end{equation}
and show in the top panel of Figure \ref{Fig1}, a histogram of $\beta$ and
$\beta'$ at $z=6$.  Without including the $E_s$ term, we find $\left <
\beta \right> = -0.295 \pm 0.0010$, and after inclusion, we find
$\left < \beta' \right> = -0.076 \pm 0.0013$.  The standard deviation
of the corrected distribution increases slightly, from $\sigma_\beta = 0.097$ to
$\sigma_{\beta'} = 0.126$.  With the inclusion of $E_s$, we find that the
virial theorem is nearly satisfied by our haloes at $z=6$:
approximately $84 \%$ lie within the cut applied by \citet{Shaw06} of
$\beta' > -0.2$, whereas only $15 \%$ of the haloes
have $\beta > -0.2$.

At higher redshifts, however, $\left<\beta'\right>$ becomes
increasingly negative, implying that our haloes are farther from
virialization at higher redshift even after the correction arising from 
the surface term is taken into account.  This is to be expected, as these
haloes have had less time to fully relax.  We show in the bottom panel
of Figure \ref{Fig1} $\left<\beta \right>$ and $\left<\beta'\right>$
as a function of redshift.  We include as the solid curve the fitting
function of \citet{Hetz06}\footnote{Note that there is a typo in their
fitting function as reported in their paper.  It should read $\eta(z)
= -3.3 \times 10^{-2} (z-2.7)^2 + 1.3$, where $\eta = 2T/|U|$.} which
they found for redshifts less than $3$. We find that the correction
term, $E_s$ does not fully correct the virial ratio at any redshift,
and the in fact correction is less effective at higher redshifts as seen in
Figure \ref{Fig1}.  


\subsection{Correlation of $\beta'$ with halo properties}

After including the extra term, $E_s$ from the virial
equation, we have found that the haloes still do not have a mean
$\beta'$ of zero as expected.  We compare $\beta'$ to other halo
properties, in an attempt to find the source of the offset.

The effect of the local environment is first examined, as it addresses
the assumption of isolation in the virial theorem. In a recent paper
\citet{Davis10} studied correlations between the local environment and
halo structural properties.  We use one of their metrics of the local
environment, the distance to the $3^{\rm{rd}}$ nearest neighbor (D3),
and compare it with the virial ratios, $\beta$ and $\beta'$. We expect
that haloes in denser environments and with close neighbors will have
larger surface terms, as these haloes are least likely to be in
isolation as is assumed by the standard virial theorem.  Thus we
expect a trend with $\beta$ and environment, and after accounting for
the surface terms, any remaining correlation could explain the
increased dispersion seen in $\beta'$ compared to $\beta$.

\begin{figure*}
\includegraphics[scale=0.6]{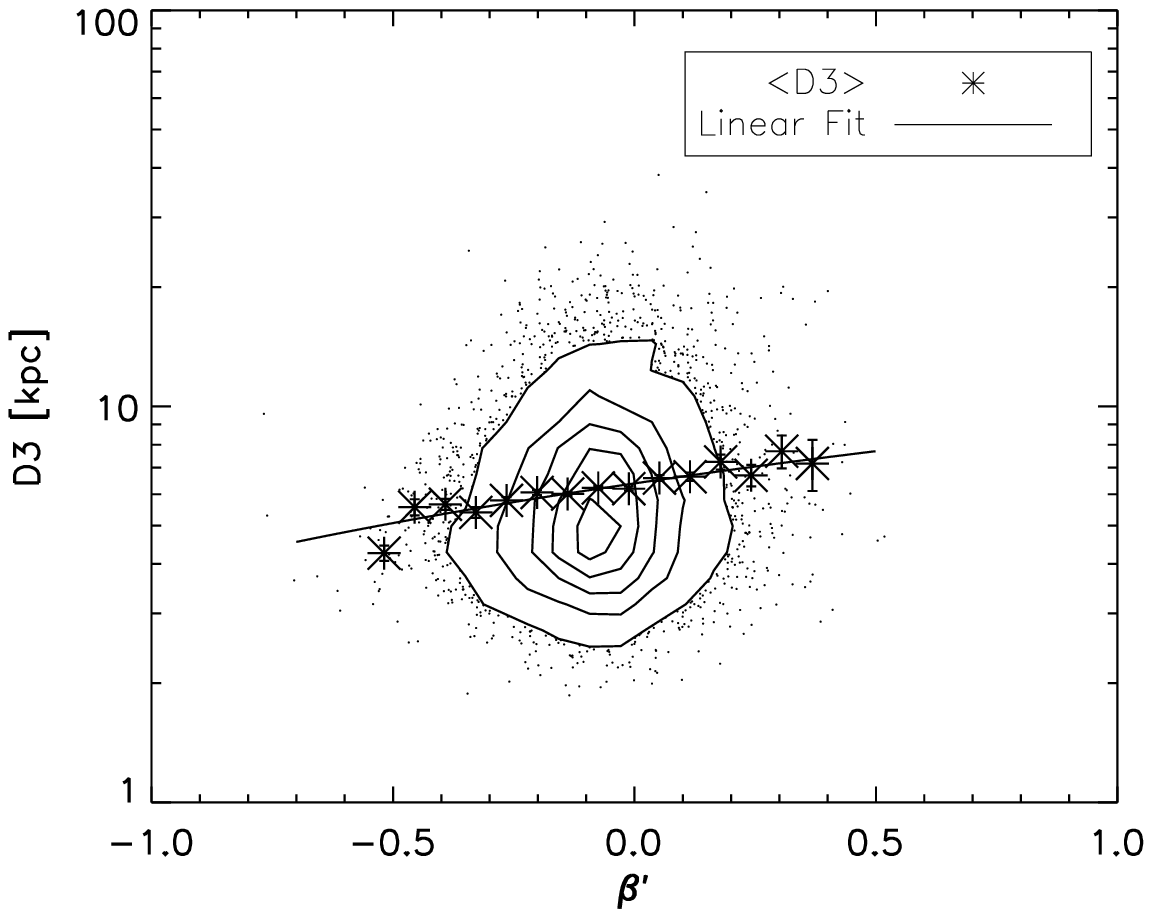}
\includegraphics[scale=0.6]{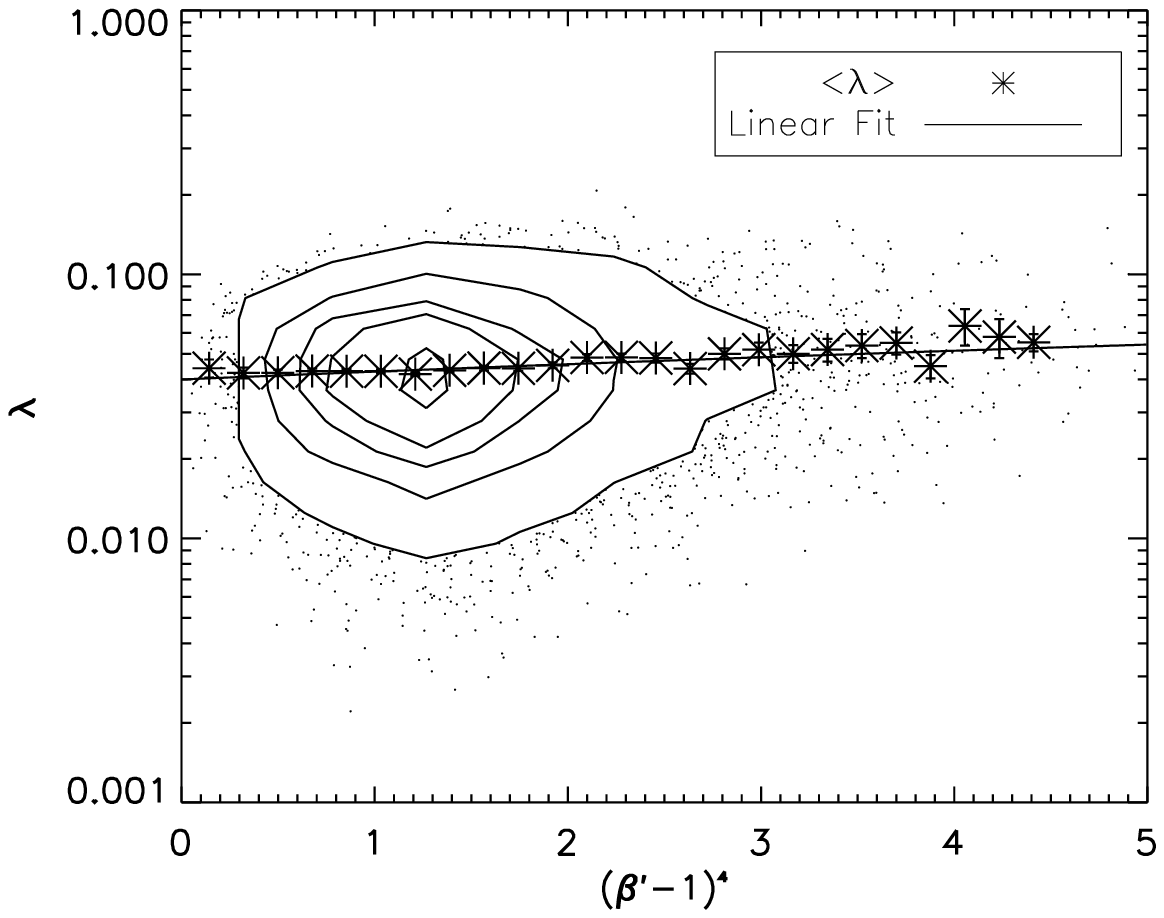}
\includegraphics[scale=0.6]{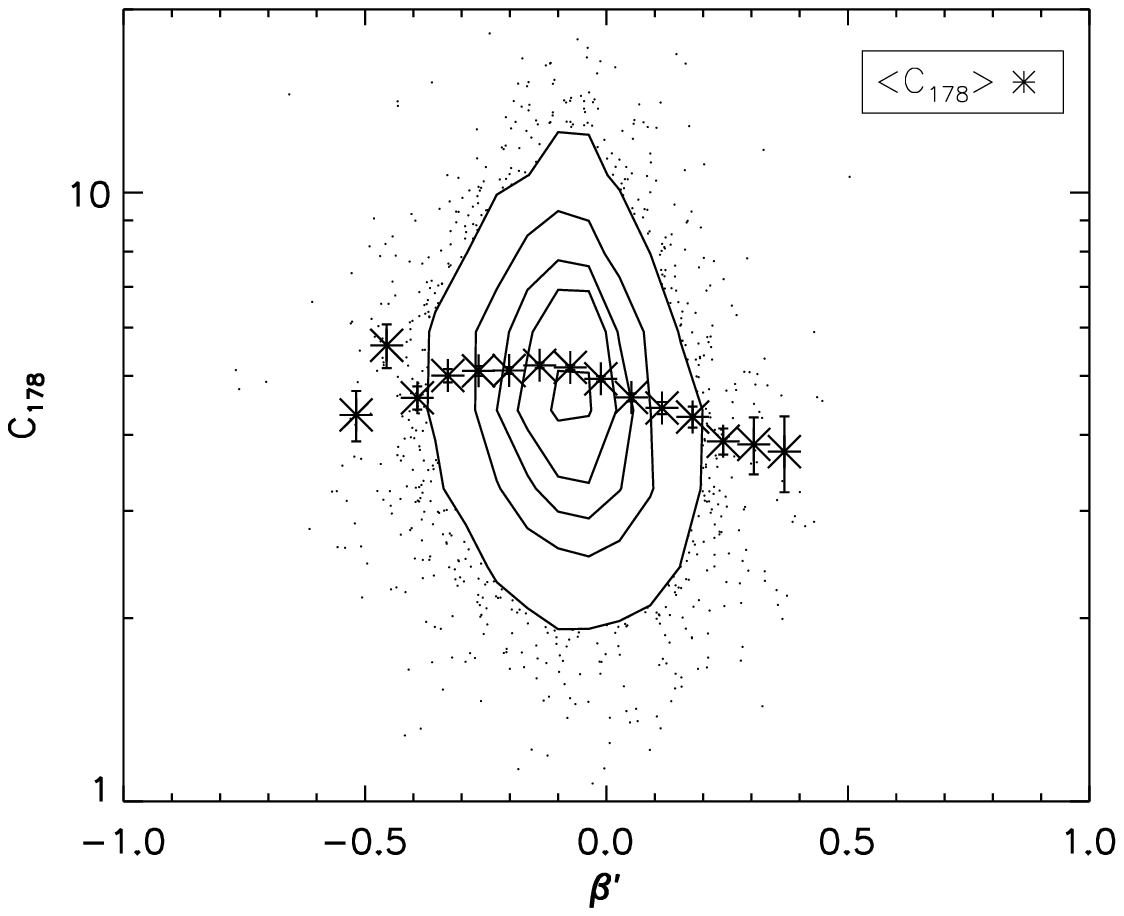}
\includegraphics[scale=0.6]{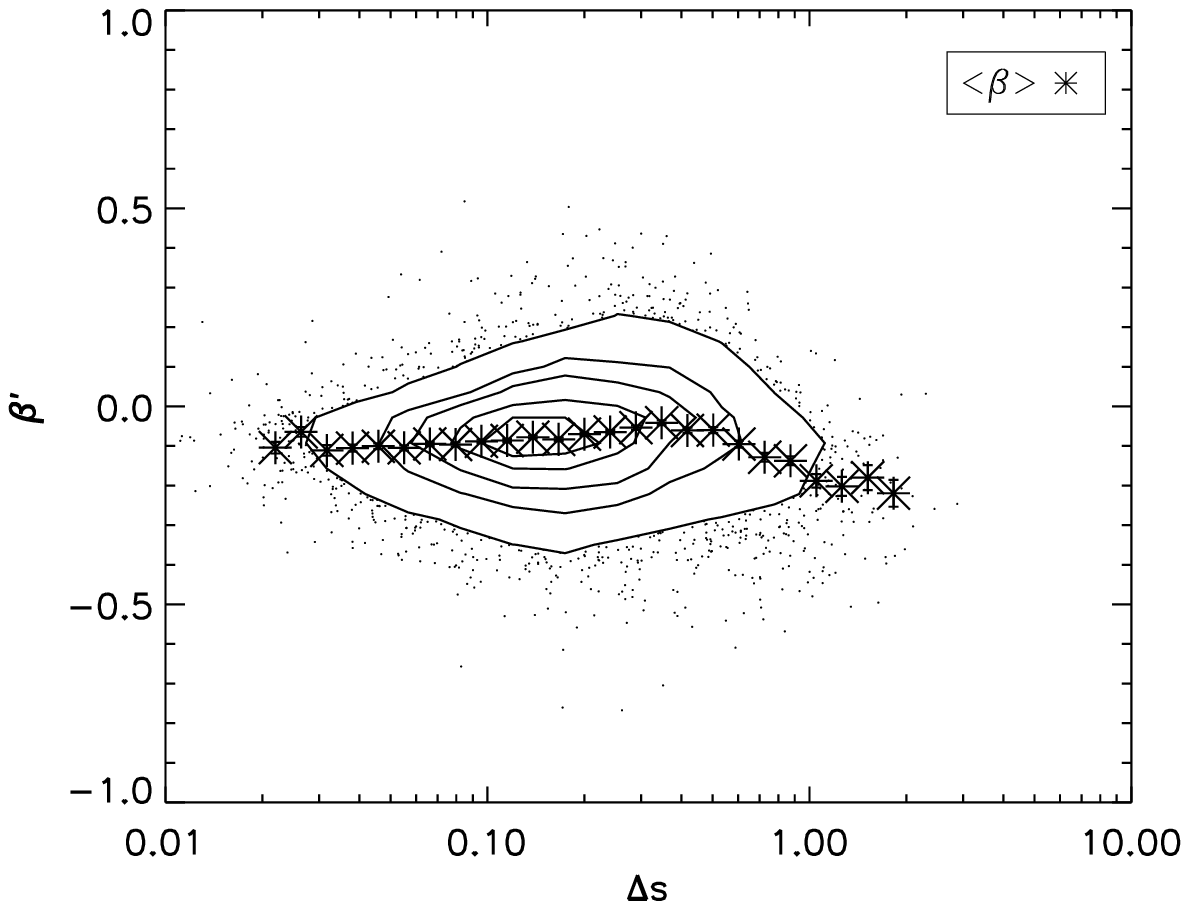}
\caption{Correlations of distance to third nearest neighbor (D3, top left panel), spin parameter ( 
top right panel, plotted versus $(\beta'-1)^4$ for comparison with
\citet{Hetz06}), concentration ($\Cs$, bottom left panel), and offset parameter ($\Delta s$, bottom right) with 
the corrected virial ratio, $\beta'$, for the haloes in our sample at $z=6$.  The contours enclose 
$90\%, 70\%, 50\%, 30\%$ and $10\%$ of the haloes, and the asterisks represent mean values for each
halo property.  We find that haloes with closer neighbors, large spin parameters,
and extremely large offsets are more likely to be further from virialization.  We do not
find the general trend reported in \citet{Shapiro04} for concentration versus $\beta'$.}
\label{Fig2}
\end{figure*}

We find that for extremely unvirialized haloes such that $\beta <
-0.45$, the mean value of D3 is $5.25~ \kpc$ with a standard
deviation of $\sigma = 1.92~\kpc$.  We note here that our box size at this
epoch is $\sim\,580$ kpc. However, for the full sample of
haloes, the mean distance is $6.19 ~\kpc$ with $\sigma = 2.94~ \kpc$.
Thus, the extremely unvirialized haloes are slightly systematically closer to
other haloes.  In Figure \ref{Fig2} we show the correlation
between $\beta'$ and D3 at $z=6$; the solid curve shows the mean value of
D3 binned by $\beta'$ with error bars depicting the $1\sigma$ error
of the mean, and the dashed curve shows a linear fit to the mean trend
with a slope $m = 2.51$ and y-intercept $b = 6.39$.  The fit has a
reduced $\chi^2 = 1.78$.  There is only a slight trend for haloes
with small values of $\beta'$ to have closer neighbors than halos with
large values of $\beta'$.  However, there is large scatter in the 
values of D3 in a given range of $\beta'$, implying that environment is not the 
only factor in the dispersion of $\beta'$.  

\citet{Hetz06} report a correlation between the spin
parameter, $$\lambda = \frac{J|T+U|^{1/2}}{GM^{5/2}},$$ and their
definition of the virial coefficient, $\eta = 2T/|U|$.  They found
that for redshifts between $0 - 3$, $\lambda$ is proportional to
$\eta^4$, and that the relationship varied with redshift.  As they
note, we expect a relationship between $\lambda$ and $\beta'$, with
scatter due to the angular momentum, $J$, in the definition of
$\lambda$.  When we use their definition of $\eta$, we find
similar results, but this definition does not account for the surface
term of equation \ref{eqn:betaprime}.  We show in Figure
\ref{Fig2} the correlation between $\lambda$ and $\beta'$ at $z=6$.  We
report a similar trend to \citet{Hetz06}.  For comparison, we fit our
data to the same relationship: $\lambda = \alpha + \gamma
(\beta'-1)^4$, and found best fit values of $\alpha = 0.04$ and
$\gamma = 2.6 \times 10^{-3}$, similar to the values
\citet{Hetz06} found of $\alpha = 0.036$ and $\gamma = 2.4 \times
10^{-3}$ for all haloes in their simulation between $0 < z < 3$.

In \citet{Davis10} we fit NFW profiles to our halo sample at $z=6$.  Using the 
scale radius from the fit, we define the halo concentration, $\Cs = \Rv / r_s$.  
In this work, we look for a correlation between halo concentration
and $\beta'$. Analytic models of dark matter haloes with galaxy and
cluster scale masses at low redshift and which use the NFW density
profile find that haloes with smaller concentrations are further from
virialization ($\beta < -0.2$) \citet{Lokas01}.  For these massive
haloes, simulations run by \citet{Shapiro04} match the analytic trends
predicted by \citet{Lokas01}.  We find, however, that our halo sample
does not follow the same trend reported in \citet{Lokas01}.  Our
simulations show that haloes which are virialized ($\beta > 0$) have
smaller concentrations than haloes with $\beta < -0.2$.

Another method of finding relaxed haloes is to use the offset between
the center of mass and the center of the potential well \citep[see
e.g.][]{D'Onghia07}. We use the location of the densest particle in the halo as a
proxy for the center of the halo's potential well, and compare this offset,
$\Delta s = (\vec{x}_{\rm{CM}}-\vec{x}_{\rm{den}})/\Rv$, to $\beta'$.
We find a small trend for haloes with small values of $\beta'$ to have
larger offsets, as shown in Figure \ref{Fig2} for the haloes at $z=6$.  However, we note
that there does not seem to be a systematic shift in $\beta'$ for
haloes with $\Delta s < 0.1$.  Thus, we conclude that using $\Delta s$
to find virialized haloes is not effective at these redshifts.

\section{Discussion}
\label{sec:discussion}

In this paper, we explore the question of whether or not high
redshift dark matter haloes are virialized. Including the surface term
in the virial equation, $82\%$ of our haloes at $z=6$ are virialized ($\beta' > -0.2$)
whereas without the term, only $15\%$ of them are.  However, the
distribution of virial ratios is still not centered around $0$. A mean of $0$
would be expected if the spread of $\beta'$ is due simply to the fact
that our measurements of the kinetic and potential energies are not
time-averaged.  Therefore, we conclude that simulated dark matter
haloes at high redshifts are not virialized. On average, they have
too much kinetic energy. One explanation for this excess kinetic
energy is due to the approximations that we have made in calculating
the terms in the virial theorem. A key issue is the choice of the
shell in calculating $E_s$, which is arbitrary.  It needs to be large
enough so that we can sufficiently resolve the flux through the
shell. However, with better particle resolution, the shell could be narrowed
for a better approximation to the surface integral.  We have also
fixed the shell as spherical, whereas the haloes are not truly
spherical.  Also, because of the time resolution of our snapshots, we
are not able to test if the moment of inertia flux (the last term in equation \ref{EQ:scalarvir}) is truly
negligible.

Our findings could be a consequence of the deployed halo finding
algorithms at high redshifts.  Since virialization happens from the
inside out, it is possible to find a virialized core by looking at regions interior
to the standard definition of the virial radius. This would be necessary if
one seeks to find fully virialized sets of particles at high redshift.
Altering the halo finder would however, alter the mass function
derived from our simulations, which would put it out of line with mass
functions from other simulations at similar redshifts \citep{reed07} 
and from analytic predictions.

Finally, our finding has important implications for galaxy formation.
Haloes with excess kinetic energy will mostly likely also have
increased turbulence when baryonic gas falls into the central regions
of the halo.  The ensuing turbulent viscosity could be an effective
method for transporting angular momentum in high redshift dark matter
haloes. This might alter the cooling and collapse of baryons in these
high redshift haloes than expected, as suggested in recent results
reported by Grief et al. (2011).

\section*{Acknowledgements}
We are grateful for the helpful comments of the referee.  We also thank Laurie
Shaw for many helpful discussions.  This work was supported in part by the
facilities and staff of the Yale University Faculty of Arts and Sciences Hight
Performance Computing Center.


\begin{thebibliography}{}

\bibitem[\protect\citeauthoryear{{Ballesteros-Paredes}}{{Ballesteros-Paredes}}%
{2006}]{BP06}
{Ballesteros-Paredes} J.,  2006, MNRAS, 372, 443

\bibitem[\protect\citeauthoryear{{Bett}, {Eke}, {Frenk}, {Jenkins}, {Helly} \&
  {Navarro}}{{Bett} et~al.}{2007}]{Bett07}
{Bett} P.,  {Eke} V.,  {Frenk} C.~S.,  {Jenkins} A.,  {Helly} J.,    {Navarro}
  J.,  2007, MNRAS, 376, 215

\bibitem[\protect\citeauthoryear{{Binney} \& {Tremaine}}{{Binney} \&
  {Tremaine}}{1987}]{BTbook}
{Binney} J.,  {Tremaine} S.,  1987, {Galactic dynamics}.
Princeton, NJ, Princeton University Press, 1987, 747 p.

\bibitem[\protect\citeauthoryear{{Bond}, {Cole}, {Efstathiou} \&
  {Kaiser}}{{Bond} et~al.}{1991}]{Bond91}
{Bond} J.~R.,  {Cole} S.,  {Efstathiou} G.,    {Kaiser} N.,  1991, ApJ, 379,
  440

\bibitem[\protect\citeauthoryear{{Cunha} \& {Evrard}}{{Cunha} \&
  {Evrard}}{2010}]{Cunha10}
{Cunha} C.~E.,  {Evrard} A.~E.,  2010, Phys Rev D, 81, 083509

\bibitem[\protect\citeauthoryear{{Davis} \& {Natarajan}}{{Davis} \&
  {Natarajan}}{2010}]{Davis10}
{Davis} A.~J.,  {Natarajan} P.,  2010, MNRAS, 407, 691

\bibitem[\protect\citeauthoryear{{D'Onghia} \& {Navarro}}{{D'Onghia} \&
  {Navarro}}{2007}]{D'Onghia07}
{D'Onghia} E.,  {Navarro} J.~F.,  2007, MNRAS, 380, L58

\bibitem[\protect\citeauthoryear{{Dunkley}, {Komatsu}, {Nolta}, {Spergel},
  {Larson}, {Hinshaw}, {Page}, {Bennett}, {Gold}, {Jarosik}, {Weiland},
  {Halpern}, {Hill}, {Kogut}, {Limon}, {Meyer}, {Tucker}, {Wollack} \&
  {Wright}}{{Dunkley} et~al.}{2009}]{wmap5}
{Dunkley} J.,  {Komatsu} E.,  {Nolta} M.~R.,  {Spergel} D.~N.,  {Larson} D.,
  {Hinshaw} G.,  {Page} L.,  {Bennett} C.~L.,  {Gold} B.,  {Jarosik} N.,
  {Weiland} J.~L.,  {Halpern} M.,  {Hill} R.~S.,  {Kogut} A.,  {Limon} M.,
  {Meyer} S.~S.,  {Tucker} G.~S.,  {Wollack} E.,    {Wright} E.~L.,  2009,
  ApJS, 180, 306

\bibitem[\protect\citeauthoryear{{Eisenstein} \& {Hut}}{{Eisenstein} \&
  {Hut}}{1998}]{HOP}
{Eisenstein} D.~J.,  {Hut} P.,  1998, ApJ, 498, 137

\bibitem[\protect\citeauthoryear{{Grief}, {White}, {Klessen} \& {Springel}}{2011}]{Grief11}
{Grief}, T., {White}, S., {Klessen}, R., {Springel}, V., 2011, ArXiv Astrophysics e-prints 1101.5493

\bibitem[\protect\citeauthoryear{{Haiman}, {Mohr} \& {Holder}}{{Haiman}
  et~al.}{2001}]{Haiman01}
{Haiman} Z.,  {Mohr} J.~J.,    {Holder} G.~P.,  2001, ApJ, 553, 545

\bibitem[\protect\citeauthoryear{{Hetznecker} \& {Burkert}}{{Hetznecker} \&
  {Burkert}}{2006}]{Hetz06}
{Hetznecker} H.,  {Burkert} A.,  2006, MNRAS, 370, 1905

\bibitem[\protect\citeauthoryear{{Jang-Condell} \& {Hernquist}}{{Jang-Condell}
  \& {Hernquist}}{2001}]{JCH01}
{Jang-Condell} H.,  {Hernquist} L.,  2001, ApJ, 548, 68

\bibitem[\protect\citeauthoryear{{Lokas} \& {Mamon}}{{Lokas} \&
  {Mamon}}{2001}]{Lokas01}
{Lokas} E.~L.,  {Mamon} G.~A.,  2001, MNRAS, 321, 155

\bibitem[\protect\citeauthoryear{{Neto}, {Gao}, {Bett}, {Cole}, {Navarro},
  {Frenk}, {White}, {Springel} \& {Jenkins}}{{Neto} et~al.}{2007}]{Neto07}
{Neto} A.~F.,  {Gao} L.,  {Bett} P.,  {Cole} S.,  {Navarro} J.~F.,  {Frenk}
  C.~S.,  {White} S.~D.~M.,  {Springel} V.,    {Jenkins} A.,  2007, MNRAS, 381,
  1450

\bibitem[\protect\citeauthoryear{{Press} \& {Schechter}}{{Press} \&
  {Schechter}}{1974}]{PS}
{Press} W.~H.,  {Schechter} P.,  1974, ApJ, 187, 425

\bibitem[\protect\citeauthoryear{{Reed}, {Bower}, {Frenk}, {Jenkins} \&
  {Theuns}}{{Reed} et~al.}{2007}]{reed07}
{Reed} D.~S.,  {Bower} R.,  {Frenk} C.~S.,  {Jenkins} A.,    {Theuns} T.,
  2007, MNRAS, 374, 2

\bibitem[\protect\citeauthoryear{{Shapiro}, {Iliev}, {Martel}, {Ahn} \&
  {Alvarez}}{{Shapiro} et~al.}{2004}]{Shapiro04}
{Shapiro} P.~R.,  {Iliev} I.~T.,  {Martel} H.,  {Ahn} K.,    {Alvarez} M.~A.,
  2004, ArXiv Astrophysics e-prints 0409173

\bibitem[\protect\citeauthoryear{{Shaw}, {Weller}, {Ostriker} \& {Bode}}{{Shaw}
  et~al.}{2006}]{Shaw06}
{Shaw} L.~D.,  {Weller} J.,  {Ostriker} J.~P.,    {Bode} P.,  2006, ApJ, 646,
  815

\bibitem[\protect\citeauthoryear{{Sheth} \& {Tormen}}{{Sheth} \&
  {Tormen}}{1999}]{ST}
{Sheth} R.~K.,  {Tormen} G.,  1999, MNRAS, 308, 119

\bibitem[\protect\citeauthoryear{{Springel}}{{Springel}}{2005}]{GADGET}
{Springel} V.,  2005, MNRAS, 364, 1105

\end{thebibliography}
\end{document}